# Low-dissipating push-pull SQUID amplifier for TES detector readout

**Mikko Kiviranta**
VTT, Tietotie 3, 02150 Espoo, Finland

E-mail: Mikko.Kiviranta@vtt.fi

**Abstract.** We suggest SQUID amplifier configuration intended for millikelvin refrigerators where cooling power is limited and hence high power efficiency is desirable. The configuration operates the SQUIDs in class-AB rather than the traditional class-A. A proof-of principle demonstration at T = 4.2 K is presented.

## 1. Introduction

Some applications of SQUIDs involve signals with very large dynamic range. Of these, multiplexed arrays of cryogenic detectors [1] tend to be simultaneously limited by the cooling budget of the millikelvin refrigerator, because interconnection parasitics make it highly advantageous to locate the first SQUID amplifier as close to the detectors as possible. The constraint in the cooling budget is particularly severe in detector systems intended to be launched to space, because every nanowatt of power dissipated at the millikelvin temperature must be lifted by the refrigerator to the radiator temperature, whereby the power gets multiplied by the refrigerator efficiency, at least the Carnot efficiency. Hence, power dissipation at millikelvin temperatures contribute strongly to the radiator size, launch mass and mission cost.

The natural flux signal range of SQUIDs is limited to the $\pm\tfrac{1}{4}\Phi_0$ monotonous range. This has been traditionally extended by negative flux feedback, either locally [2, 3] or via room-temperature electronics [4] in the so-called flux-locked loop (FLL) configuration. In the local feedback, unfortunately, the output signal level is lowered so that additional SQUID amplifier stages are typically needed. In the FLL the cable delay and the requirement of the feedback loop stability limit the performance in a fundamental way.

In other hand, the natural dynamic range of the dc-SQUID, ie. the ratio of the $\Phi_0/2$ monotonous flux range to the flux noise $\Phi_N$, determines its power dissipation (see Appendix, also [5]). Ways to alleviate the power dissipation include use of L-SQUIDs [6, 7], operating dc-SQUIDs inside a long or local negative feedback loop, or applying fluxon counting techniques [8]. Here, we suggest a technique adopted from room-temperature electronics, in particular HiFi audio equipment, namely operating the active devices of the amplifier in class-B or class-AB (see [9] ch. 18-3) rather than class-A as is customary with dc-SQUIDs. This work is motivated by the planned ATHENA [10] and SPICA [11] space observatories.

## 2. Design

The push-pull circuit (fig. **1**a) is often used in room-temperature transistor amplifiers to improve power efficiency, i.e ratio of the dissipated power to the signal power delivered to the load. It takes advantage of the feature of transistors that they are almost perfect insulators in their inactive state. Even if there may be a large voltage across the transistor, the zero current implies zero dissipation. The SQUID has the dual feature of being a perfect conductor in its inactive state, with the role of currents and voltages reversed. We have demonstrated operation of a SQUID amplifier in the AB/B-class operation, using a network topology dual [12] to the transistor-based push-pull stage (fig. **1**b).

Referring to the fig. **2**, in the traditional class-A amplifier configuration the SQUID dwells statically at the point 'a', and traverses between points 'b' and 'c' at the maximal flux excitation. In our class-B configuration, in absence of flux signal both T1 and T2 dwell at the zero-dissipation point 'd'. When a sinusoidal flux excitation is applied, during the first half-cycle the T1 traverses from 'd' to 'e' and back, while T2 remains in the superconducting state. During the second half-cycle the roles of T1 and T2 are reversed. The superconducting state of T1 (T2) is indicated as the flat part of the cyan-colored flux

# Low-dissipating push-pull SQUID amplifier for TES detector readout

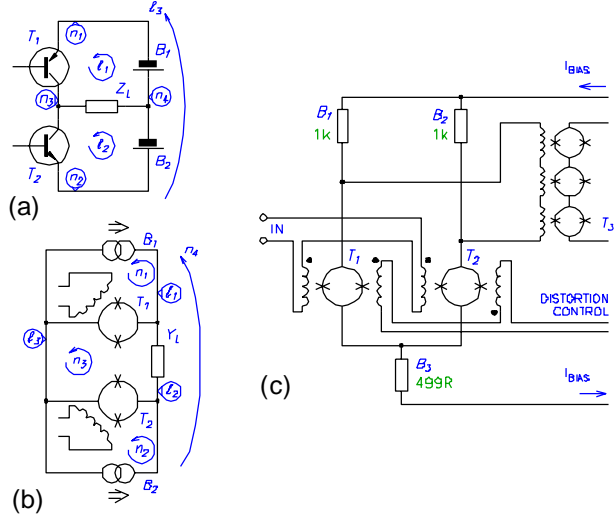

Figure 1: The dual transform swapping impedances Û admittances, currents Û voltages and nodes $n$ Û mesh loops l, has been applied to the standard push-pull transistor circuit (a). The resulting equivalent SQUID circuit (b) was implemented in practice as shown in (c).

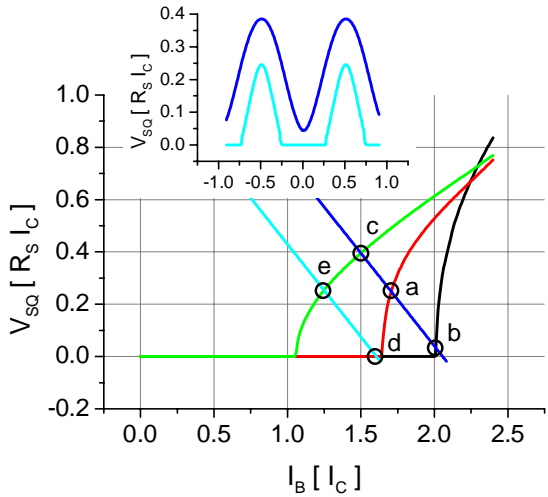

Figure 2: Typical current to voltage characteristics of a dc SQUID at $n\Phi_0$ (black), $(n + ¼)\Phi_0$ (red) and $(n + ½)\Phi_0$ (green) applied flux. Load lines for biasing from $0.7·R_S$ bias resistance at $I_B = 2.05·I_C$ (blue) for the class-A operation and $I_B = 1.6·I_C$ (cyan) for the class B operation. The insert shows the flux-to-current SQUID response along the two load lines as a function of applied flux. $I_C$ is the critical current of a Josephson junction and $R_S$ is the shunt resistance.

response in the inset of the fig. **2**. During the first half-cycle the current from the source B1 minus the current through the SQUID T1 is guided to the input coil of a high-power booster SQUID T3, and similarly for B2 and T2 during the second half-cycle. The SQUIDs T1 and T2 are to be located in the 50 mK stage and the T3 in the 2.5 K stage eg. of the X-IFU refrigerator [10].

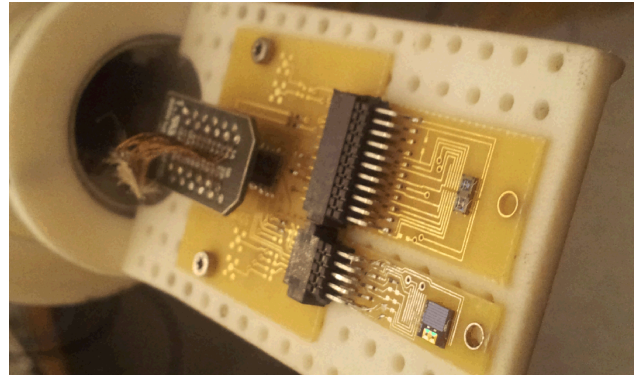

Figure 3: Experimental setup in a 4.2K dipstick. The upper card carries a dual J3 SQUID chip, and the lower card the L5 booster SQUID chip.

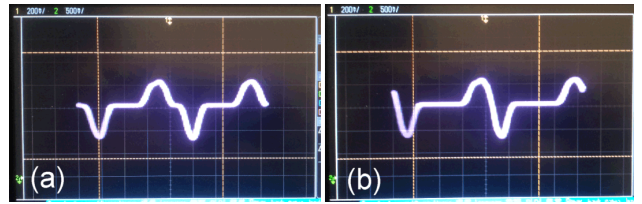

Figure 4: Oscilloscope traces of the *T3* output voltage, when the flux input 'IN' is excited and *T1* and *T2* are operated at a low bias current $\sim 1.2 \times I_C$ for clarity of traces. (a) *T1* slope above the superconducting line does not coincide with the *T2* slope below the s.c. line, and crossover distortion is present. (b) Flux is applied via the 'distortion control' line of fig. 1c to merge the *T1* and *T2* slopes into one continuous slope.

## 3. Experiments

For the demonstration purposes we realized the class-B circuit of fig. **1c** with all the SQUIDs immersed in liquid helium at 4.2 K as shown in fig. **3**. We used the J3-type small SQUID arrays as the T1 and T2, and the L5-type SQUID array as the T3. Our J3-type is a 6-series SQUID array and the L5-type is a 184-series 4-parallel SQUID array, constructed of projection lithography versions [13] of our cascadeable SQUID cells [14]. Constant current sources B1 and B2 were approximated by cold resistors. After cross-over distortion was adjusted as shown in fig. 4, the amplifier functioned as designed. In the strictly class-B setpoint where both T1 and T2 are superconductive in the idle state, the amplifier showed a lot of excess noise. Adjusting the setpoint to the class-AB where T1 and T2 are in a low-dissipating finite voltage state, the amplifier showed 0.17 $\Phi_0/Hz^{1/2}$ flux noise and 5 MHz bandwidth, see fig. **5**. Given the $M^{-1} = 36$ mA/$F_0$ mutual inductance of the J3 input coils, the input-referred current noise was about 6.5 pA/Hz$^{1/2}$.

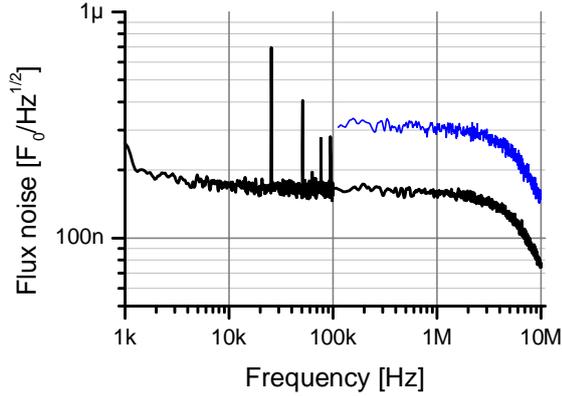

Figure 5: Measured flux noise of the class-AB amplifier. Also shown is the frequency response, measured by injecting white pseudorandom noise as the flux signal.

## 3. Discussion

### 3.1 Operation with Baseband Feedback

The class-B amplifier is particularly advantageous as a part of negative feedback schemes where the SQUID effectively acts as a null detector. Of special interest to us is operation with so-called Baseband Feedback (BBFB) [15, 16] as applied to X-ray microcalorimeters [10]. In such a configuration the class-B amplifier spends most of its time in the low-dissipating idle state, waiting for X-ray events.

Baseband feedback is an extension of the FLL applied to the special case of Frequency Domain Multiplexing [17] which utilizes amplitude modulated high-frequency carriers. The insight of the BBFB is that the carriers are deterministic and hence can be deterministically reconstructed, and only the low-bandwidth modulating envelope requires feedback.

The static power dissipation of the circuit depends on the accuracy at which the 'distortion control' line can be adjusted between the low-noise class-AB and noisy class-B setpoint. We estimate that the static dissipation at setpoint 'd' can be reached which is 5 … 10% of the dissipation at the setpoint 'a'. The dynamic power dissipation would then depend on the duty cycle of the X-ray events. With duty cycle < 10% static dissipation would dominate.

### 3.2 Various remarks

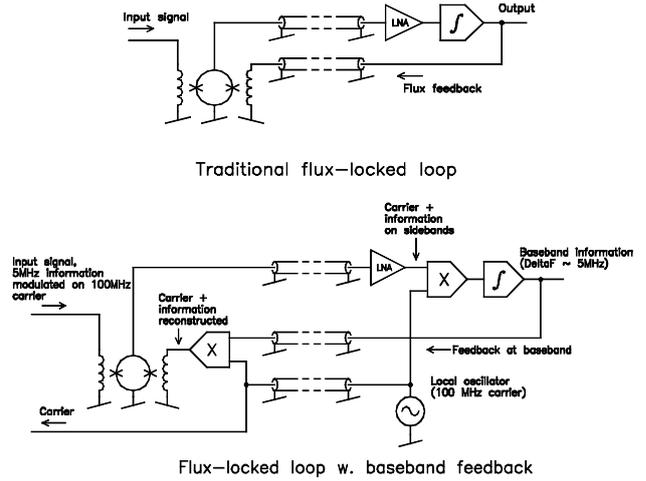

Figure 6: Baseband feedback seen as an extension of the Flux Locked Loop. Figure taken from ref. [**15**].

As the SQUIDs T1 and T2 are operated at bias current less than $2\times I_C$, they can be designed with higher McCumber parameter than the canonical $b_C = 0.7…1$ without the risk of encountering the hysteretic region. This would lead to a lower flux noise. This is counterweighted by the fact that the input inductance of the T1 + T2 combination is twice that of the single SQUID.

The observed $F_N = 0.17$ $\mu F_0$/Hz$^{1/2}$ flux noise is somewhat better than typically observed for single J3-type SQUIDs at T = 4.2 K. It would be tempting to assign the difference to the sqrt(2) averaging due to two devices present. As the second device is always inactive half of the cycle, a more likely explanation has to do with the unusually low-bias operating region for the J3 devices.

The load $Y_L$ shown in fig. **1b** in our actual experiment was reactive, i.e. inductance of the T3 input coil. Hence the load line d-e of fig. **2** was almost vertical for low frequency signals, tilting towards the attitude shown in fig. **2** at high frequencies where reactance increases.

The observed 5 MHz amplifier bandwidth was probably limited by a poorly understood problem in our dipstick wiring and its interaction with L5-type SQUIDs. From the dynamic resistance $R_D \approx 20$ Ω of the J3-type SQUIDs and the L5 input inductance $L_{IN} \approx 180$ nH plus interconnect parasitics we would expect a bandwith order-of 12 MHz.

The added wiring complexity due to the extra 'distortion control' line may be lifted if SQUIDs with intentional bias to flux coupling are used.

## 4. Conclusion

We have demonstrated low noise operation of a two-SQUID push-pull amplifier circuit at the liquid He temperature. The circuit was operated in class-AB, analogously to transistor-based push-pull circuits, and it was estimated to have a higher power efficiency than ordinary SQUID amplifiers. The input-referred flux noise of 0.17 $\mu\Phi_0/\text{Hz}^{1/2}$ was observed, which is lower than same SQUID devices show when operated in a single-SQUID class-A configuration. We expect to see 10 – 20 times lower dissipation than class-A operation in practical instrumentation when similar dynamic ranges are targeted.

**Acknowledgement**

This work has been supported by the AHEAD project, European Union's Horizon 2020 research and innovation programme, grant agreement No 654215.

**APPENDIX**

*Rule-of-thumb power dissipation of the dc-SQUID operated in class-A.* Referring to fig.2, the bias current of the point 'a' should be chosen slightly higher than $I_B = 2 \times I_C$, in order the flux swing c-a-b not to clip at the superconducting line. For a SQUID with dimensionless inductance $b_L = 1$ the point 'a' current is roughly $1.5 \times I_C$, The voltage along the matched-bias load line is $0.5 \times I_C$ times the slope $0.7 \times R_S$. Rounded upwards, the resulting power dissipation is $P_D = 0.6 \, R_S \, I_C^2$ in terms of the shunt resistance $R_S$ and one-junction critical current $I_C$.

*Connection between SQUID dissipation and dynamic range.* Flux noise of a dc SQUID is $\Phi_N = L_{SQ}\sqrt{x k_B T / R_S}$ where $x \approx 14$ for an optimal autonomous SQUID, but whose precise value is of no interest here. Substituting the definition of the dimensionless inductance $b_L = 2 L_{SQ} I_C / \Phi_0 \approx 1$, one notices that the ratio of monotonous flux range of $\Phi_0/2$ to the flux noise $\Phi_N$, i.e. amplitude dynamic range per 1 Hz bandwidth denoted as $D$, is $D = \sqrt{P_D} \cdot (0.6 x k_B T)^{-1/2}$. Here $k_B$ is the Bolzmann's constant and $T$ is electron temperature of shunt resistors, usually limited to $T > 0.3$ K due to electron-phonon decoupling.